# High Resolution Particle Beam Monitoring and Ionization Counters with the Help of Single Carbon Nanotubes

K.A. Ispirian, R.K. Ispiryan and A.T. Margarian

Yerevan Physics Institute, Brothers Alikhanian St, Yerevan, 0036, Armenia

**ABSTRACT**

After a short review of modern beam monitors, ionization and proportional counters and discussion on the necessity to have thinner wires, we propose and consider construction and parameters of nanotube particle beam monitors and counters.

**Keywords:** Carbon nanotubes, beam monitoring, particle detectors.

1.   Introduction

As it has been shown in [1], the nanocrystals of carbon nanotubes (NT) can find application in high energy and X-ray physics (see the review [2]). Films of oriented NTs already serve as cathode emitters for electron [3] and X-ray beam production [4], and there is a proposal [5] to construct particle detectors with better ability of particle identification using such films. On the basis of not oriented NTs, there are tested radiation detectors sensitive to UV [6] and IR [7]. The advance of the methods of NT production and nanomanipulation allows us to ask: can isolated single NTs be used in high energy physics due to their unique physical properties? At present long NTs, in particular, separate single wall nanotubes (SWNT) with length more than 4 cm are produced [8]. SWNT and multi wall nanotubes (MWNT) have diameters equal to a few and few tens of nm, respectively, i.e., they are much thinner than the diameter of metallic wires and the laser beam diameter which are of the order of a few microns. This means that nanotubes can provide better spatial resolution and higher electric field gradients than the best of the tested modern beam particle monitors (BPM) [9,10] and ionization/proportional counters. The Youngs modulus of single NTs is greater than that of stainless steel and quartz, they have sufficient thermal and electrical conductivity [11] and, as it has been shown in many works beginning from [12], NTs have sufficiently high radiation hardness.

Stainless steel, carbon and tungsten scanning wires having large Young's modulus are preferable for high intensity BPM with resolution of the order of the wire diameter. The horizontal and vertical dispersions, $\sigma_x$ and $\sigma_y$, of the beam cross sections at the interaction points in SLC, SLAC are equal to 1.5 and 0.65 $\mu m$, while in future ILC and CLIC they will be equal to 0.6 and 0.006 $\mu m$ and 0.043 and 0.001 $\mu m$, respectively [10]. Thinner wires are required for better profile resolution. However, as the experiments show, wires with diameters less than ~4 micron cannot be used since they undergo destructive damages due to thermal energy deposition if the rms transverse size of $10^9$ electrons is less than 1 micron (see [9,10]). To solve these problems, it is proposed and tested to replace the solid wire scanner by thin laser beam and to detect the Compton scattered photons [13] or electron-positron pairs [14]. With the help of interference methods the resolution of the recent laser BPM had achieved a few tens of nm [9].

Recently, a new method [15] of vibrating wires has been tested on beam of charged and neutral particles, the principle of operation of which briefly can be described in the following way. It is well known that each wire, fixed in its two ends, has resonance frequency of transversal oscillations which depends on material properties and length. If such an oscillating wire is irradiated by beam particles, then, due to ionization losses and energy deposition, the temperature of the wire will be increased resulting in a lengthening of the wire due to thermal expansion. Measuring accurately the new oscillation frequency, one can determine the temperature variation, and therefore the flux of the particles passing through the wire.

On the other hand, in ionization gas counters and multiwire proportional counters (MPC), the thin metallic wire anodes around which large gradient of electrical fields are created usually have thickness 20 - 100 microns [16,17]. Again the use of thinner wires can result in useful effects; however, mechanical and electrical problems do not allow anode wires with diameters less than ~10 $\mu m$.

In this article we will consider the physics, advantages and possibilities of construction of BPMs and single particle ionization counters using separate conductive NT wires. Many technical problems as well as the possibility of the more complicated use of single NT field effect transistors (FET) sensitive to radiation are postponed for further publications.

2.  **BPM with NT Wires**

As it is well known, the SWNT with helicity given by a pair of integer numbers n and m equal to each other are metallic, while the SWNT with unequal indices and almost all the MWNTs are semiconductors (see [2]). The SWNT with n=m=10, which will be considered mainly in this work, has the following properties (see [11]): diameter ~1.3 nm, density ~1.33 g cm$^{-3}$, Young's modulus ~ 1 TPa, maximum tensile strength ~30 GPa, electrical resistance, ~10$^{-4}$ $\Omega\ cm$, thermal conductivity ~2000 W/m/K and negative coefficients of thermal expansion ~6·10$^{-6}$ K$^{-1}$ [18]. To our knowledge, the physics of the radiation hardness of isolated SWNT in various particle beams has not been studied satisfactorily (for some results, see [19]). The methods of 3D manipulation of NTs under scanning electron microscope providing nm accuracy have been developed mainly for electronics (see [20,21] and references therein). Without discussing these methods we assume that one has (see Fig. 1) a 1-2 cm long NT conductive wire (1) tidily stretched by two clips (3) on a shifted support (4), made of material with low coefficient of thermal expansion (CTE). The fast scanning is made with the help of a driver (5) or of the steering magnet (7) in the case of neutral and charged particle beams, respectively. The particle beam profile is determined by measuring the interaction of particle with NT atoms in each wire position by the following methods: 1) Detecting the secondary particles under small angles as it is shown by the detectors (6); 2) Recording the electric current induced in the NT wire due to secondary electron emission as in [22]; 3) Measuring the frequency of the vibrating wire as it is described in [15] or 4) Detecting the ~200 eV characteristic radiation of carbon atoms as in [23].

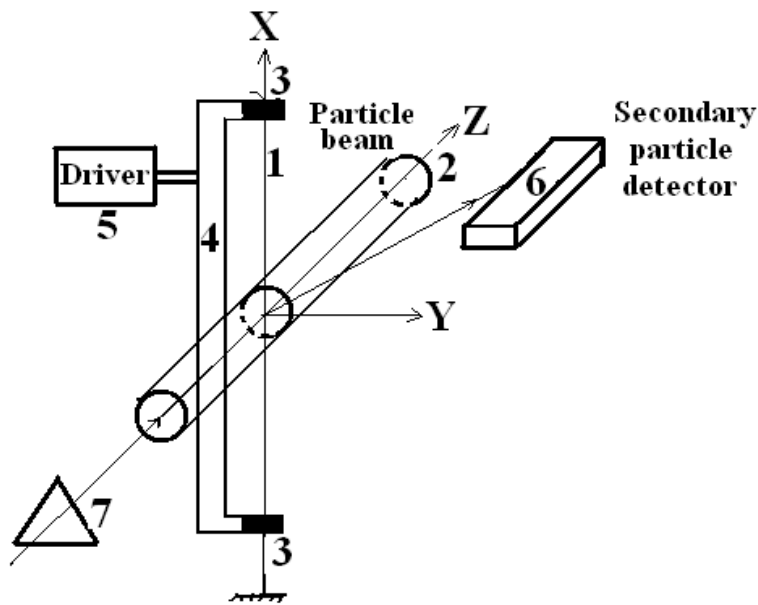

Fig.1. A stretched SWNT for BPM  1) CN; 2) a particle beam; 3) clipses; 4) wire support; 5) driver; 6) secondary or scattered particle detector; 7) steering magnet.

Let us make some simple estimates on mechanical and thermal properties of the NT wire. Let $l_0$ be wire length without any stress and without any radiation at the initial temperature $T \approx 300\,K$, and the clipses giving additional stress to NT make the length of the strained wires $l_B$, so that the initial stress of NT is equal to $\sigma_0 = E(l_B - l_0)/l_0$ which we keep greater than about half or much less than the maximum tensile strength equal to ~30 GPa. It seems reasonable to expect that, as in the case of metallic wires, the temperature of the NT wires under irradiation will be increased due to the energy deposited in NT, which in some approximations can be estimated (see, e.g. [24]). However, as it will be shown below, the physical processes of energy deposition are quite different and the temperature of the NT wire can not be increased higher than ~600 K ($\Delta T^{NT} \approx 300\,K$). Taking into account that the support and NT have coefficients of thermal expansion (CTE) of different sign (the stresses of NT and of support are summed), but the temperature of the support will not be changed ($\Delta T^{sup} \approx 0$) the tensile strength of NT during the radiation will be equal to $\sigma_T^{NT} = [E^{NT}(CTE)^{NT}\Delta T^{NT} + E^{sup}(CTE)^{sup}\Delta T^{sup} \approx E^{NT}(CNT)^{NT}\Delta T^{NT}$, which is much less than 30 GPa. Therefore, NT will not be damaged also under radiation. Since the convenient for the measurements the second harmonics frequency of the wire oscillations is equal to $F = 1/l_0 \sqrt{\sigma_0/\rho}$ =0.4246 MHz, the variation of the second harmonic frequency is equal to $\Delta F/F = (E/2\sigma)(\Delta l_T/l_0) = (E/2\sigma)(CTE)\Delta T^{NT}$ and is much larger than that for metals, because for NT the coefficient $E/2\sigma$ is greater. Therefore, one can apply the vibrating wire method [15] with NT too.

Now let us consider the processes resulting in energy deposition in NT wire which differ essentially from those which take place in metallic wires. Indeed, still in [25] it has been shown that the distribution of the ionization energy losses in thin layers differs from that predicted by the Landau theory [26] or other models [27], and in the case of very thin layers, when the probability of collision of beam particles with layer's atoms is much less than unity, this distribution has discrete forms peaked at certain energies. Having these losses and the probabilities of absorption of the emitted soft and hard photons as well as of $\delta$-electrons, in principle, one can calculate the energy deposited in the layer. Such calculations taking into account restricted energy losses due to escape of $\delta$-rays are given, for instance, in [28]. However, just as graphen, the single NTs present two or a few bent layers of carbon atoms, and to our knowledge, there are no such calculations on interaction rates and energy deposition in isolated NTs. It is clear that only a negligible fraction of the produced secondary electrons or photons produced by ionizations or excitations are absorbed by the few carbon layers, and the probability of nuclear interactions of passing beam particles with carbon nuclei resulting in structure defects is very small. Therefore, the expected energy depositions due to passing beam energy losses and increase of the wire temperature must be very small. With this, the above assumed maximal temperature less than 600 K of CNT during irradiation seems true. There is no doubt that the calculation of the realistic increase of the NT temperature is a very hard problem, however, as it has been shown above, it can be measured by the vibrating wire method.

Considering the rate of the BPM measurements of four methods shortly described above we shall make estimates only for the newly proposed 4) characteristic radiation method since the other three are connected with relatively larger cross sections and are widely used. We shall assume that the particle beam bunch has square cross section $S_{Bunch}=L_x L_y$ and contains $N_{Bunch}$ particles with uniform density distribution instead of usual Gaussian distributions. In the case of isolated SWNT target, when the particle beam passes only $n = 2$ atomic layers, and the beam cross section is larger than the NT's transversal size, it is reasonable to estimate the number of the produced characteristic quanta in a frame where the beam particles are at rest using invariance properties of certain physical quantities. One can show that the probability of interaction of single target atom with the beam bunch containing $N_{Bunch}$ particles is equal to $P = n_{Bx}\sigma_{int}$, where $n_{Bx} = N_{Bunch}/S_{Bunch}$ is the surface density of the beam particles "seen" by passing NT atoms, and $\sigma_{int} = \sigma_K$ is the interaction (in our case the K-shell ionization) total cross section. Therefore, the number of the detected characteristic radiation photons per bunch is equal to $N_{Bunch}^{DetK} = F N^{NT} \Delta\Omega_{Det} N_{Bunch} \sigma_K / 4\pi S_{Bunch}$, where $F$ is the target NT atom fluorescence yield, $N^{NT}$ is the number of an atoms of 2 layers of an NT which can interact with the beam cross section, $\Delta\Omega_{Det}$ the acceptance solid angle of the detector.

The cross section of K-shell ionization is equal to [29,30]

$$\sigma_K = \sigma_0 \left| \frac{1.5}{I} + \frac{0.423}{I} \left| \ln\left(\frac{1.274}{I}\gamma^2\right) - 1 \right| \right|, \qquad (1)$$

where $\sigma_0 = 8\pi r_e^2/3 = 0.665 \times 10^{-24} cm^2$ and $I = E_K/mc^2$ is the K-electron binding energy in units of electron mass energy. As it follows, the K-shell cross section grows logarithmically with the increase of energy without revealing density effect as in case of ionization energy losses, because harder photons give contribution into K-shell ionization. The expression (1) is in good agreement with experimental data [31] up to electron energies 900 MeV, and one can expect it will be correct also in our case at higher energies since the density of the carbon atoms in NT is very low.

In the case of ILC with $L_x = 6 \times 10^{-5} cm$, $L_y = 6 \times 10^{-7} cm$, $E_e = 500\, GeV$ or $\gamma = 10^6$, $N_{Bunch} = 10^{10}$, $n_{Bunch} = 2.8 \times 10^{20}\, cm^{-2}$, for NT

SWNT(10,10) $I=4\times10^{-4}$, $N^{NT}=600$, fluorescence yield $F=0.002$, $\sigma_K=2.6\times10^{-20}$ $cm^2$ and for $\Delta\Omega=10^{-2} rad$ one obtains $N_{Bunch}^{Det,K} = 6.7\times10^{-3}$. Taking into account that for ILC the number of bunches per second is ~$1.25\times10^4$, about 80 characteristic photons per second will be detected for the above parameters from a NT BPM on ILC. Such a quantity of characteristic photons is enough to carry out the measurement of the beam profile.

3. **Proportional Counters with NT Wires**

Now let us consider the construction and principle of operation of a proportional counter with CNT anode schematically shown in Fig. 2. In ionization counters the main requirement on the anode wire is to provide smooth and stable electrostatic field in the space between anode and cathode, especially near the anodes, where the main processes take place. This means that the wires should have smooth surfaces. However, the production technique of thin wires does not exclude roughness of the sizes of hundred nm. From this point of view CNTs surfaces containing hexagons of carbon atoms are sufficiently good, because they provide smooth electrostatic potentials at distances from CNT surface larger than a few times maximal distance between the carbon atoms (0.28 nm). CNT has also the advantage of stability and is chemically less reactive.

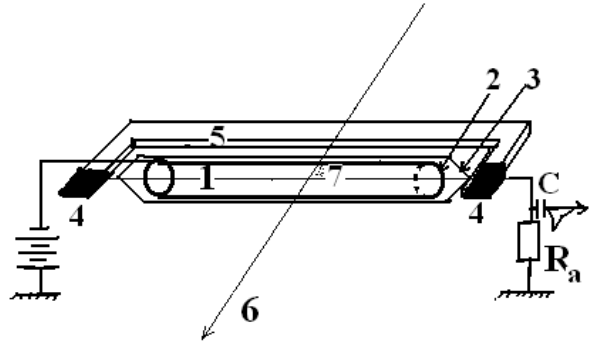

Fig. 2. A proportional chamber. 1) CNT with $r_a$ = 1.3 - 50 nm anode; 2) cylindrical metallic cathode with $r_c$ = 5 - 12 mm; 3) glass tube containing appropriate gas; 4) clipses of 5) CNT support. 6) is a particle producing the electron-ion shower 7).

Since the electric field $E(r)$ at the point $r$ near the anode of a cylindrical counter with cathode and anode radii $r_c$ and $r_a$ and applied voltage $V_0$ is given by the expression [16,17] $|\vec{E}(r)| = V_0/[(r \ln r_c/r_a)]$, the replacement of a metallic wire with diameter $r_a$ = 20 $\mu m$ by CNT with $r_a \sim$ 20 nm for fixed other parameters results in increase of the field ~10 times everywhere, and provides an additional space of 20 μm for gas multiplication, where the maximal available field is increased toward to the cathode surface ~1000 times. In such high gradient fields the shower produced by charged particles in the counter will be amplified. The amplification factors $G$ for proportional chambers have been calculated in various models in various approximations and compared with available experimental data [32-38]. In the recent works $G$ is evaluated by Monte Carlo simulations [37]. In [38] in particular it has been shown experimentally and by simulations that the proportional counter's resolution is better for thinner anode wires. Unfortunately, there are no theoretical and experimental studies devoted to dependence of $G$ on the anode radius. For the purpose of obtaining such dependences, it is enough to estimate $G$ using [35] since, as it has been shown experimentally in [39], the model [35] provides sufficiently correct estimates. Besides, the predictions of the model [35] do not depend on a threshold voltage, which for majority of other methods can be determined only from experimental data. The model [33] also has this last advantage; however, it is applicable for lower values of electric field.

According to [35]:

$$\ln G / \Pr_a S_a = A(1 + CS_a)\exp(-B/S_a),\qquad(2)$$

where $P$ is the gas pressure, $S_a = E_a/P = V_0/[\Pr_a \ln(r_c/r_a)]$ is ratio of the electric field (at the anode surface) and gas pressure. $A$, $B$ and $C$ are constants given in [35] for various gas mixture and pressures. Compared with other models the model [35] is applicable for higher values of $S_a$.

Fig. 3 shows the calculated dependence of the amplification factor on the anode radius for a counter with $r_c$ = 6 mm filled with xenon, for 3 values of applied voltage $V_0$ (Fig. 3a) and of gas pressure $P$.

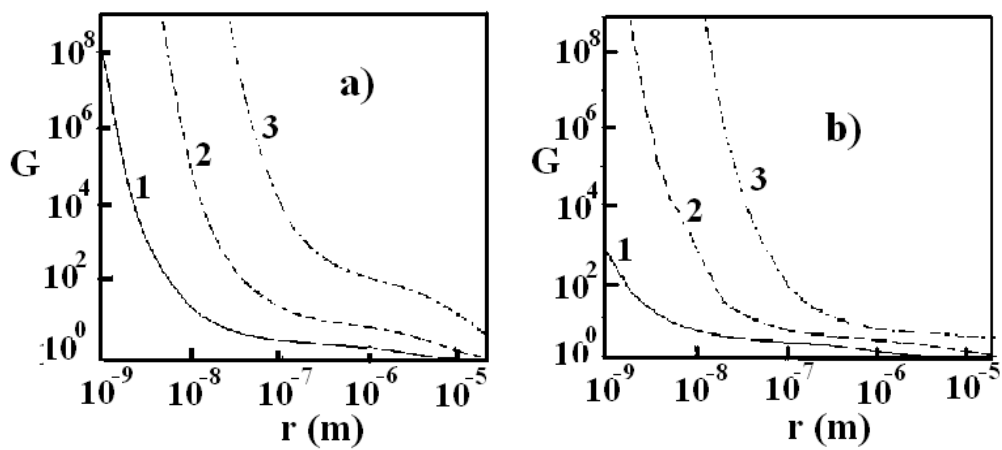

Fig.3. The dependence of $G$ upon $r_c$ for $V_0$ = 600, 300 and 150 V (curves 1, 2 and 3) a) and for $P$ =0.1, 0.5 and 3.0 atm (curves 1, 2, and 3) b) at other fixed parameters (see the text).

It is expected that the predictions of the model [35] are valid for values of $S_a$ ~ 5000 or of $G$ ~ $10^5$. As it follows from the curves, the proportional counters with CNT wires provide higher $G$ for lower values of $V_0$ and $P$. Such properties can be important for avoiding the space charge effects in the case of operation with $\alpha$-particles and fission products and justify the difficulties of CNT application.

## 4. Conclusion and Discussion

Since the diameter of CNT is slightly larger or equal to a few nm, it is not reasonable to use them as BPM for machines for which the beam radius is greater than 1 $\mu m$ (LCLS, Tevatron, LCLS, etc), and the tested methods of non-intercepting (Cavity BPM) and intercepting (wire scanners) can provide satisfactory results. CNT are of interest only for SLC ($\sigma_x$ =1500 and $\sigma_y$ =650 nm), ILC ($\sigma_x$ =600 and $\sigma_y$ =6 nm) and CLIC ($\sigma_x$ = 43 and $\sigma_y$ = 1 nm) with their electron and gamma quanta beams.

The most difficult problems facing the proposed CNT method are connected with the facts that it is not possible to see CNTs, and that the accuracies of their aligning and surveying for long term are less than hundreds of nm. Their discussion is postponed for future advancements in nanotubes' manipulation.

Since in the case of using a CNT anode instead of ~micron diameter wires the electric field gradients are much higher, it is necessary to develop a new theory or model approximating the ionization cross section and Townsend first coefficients for higher fields. It is convenient to carry out the experimental test of the above predictions in more complicated low pressure MWPC in which one of the metallic wires is replaced by CNT because in such a device one can expect to carry out reasonable comparison of the pulses obtained from thicker wires and thinner CNT wire at similar other parameters. One can expect that in MWPC in which all the wires are replaced by CNT mutually perpendicular planes, it will be possible to achieve better spatial resolutions that a few tens of microns achieved in modern silicon strip vertex detectors.

Finally, it is necessary to mention that the use of semiconductor CNT as particle detector devices with n-p or p-n transitions in BPM or proportional counters will results in higher spatial resolution. Just as nanotubes will replace the silicon technology in electronics and other fields, the future particle detection and identification will be dominantly based on nanotubes. However this will depend on the advances in nanotube electronics [40].

**References**

1) L.A.Gevorgian, K.A. Ispirian and R.K. Ispirian, *Nucl. Instr. and Meth. B*, **145**, 155, 1998.

2) X. Artru, S.P.Fomin, K.A. Ispirian, N.P. Shulga, N.K. Zhevago, *Phys. Rep.* **412**, 89-189, 2005.

3) Y. Houmi, S. Ohsawa, T. Sugimura, M. Ikeda, Proc. LINAC04, Lubeck, Germany, 2004.

4) G.Z. Yue et al, Appl. Phys. Lett. 81, 355, 2002.

5) K.A. Ispirian, R.K. Ispiryan, Proc. SPIE, 6634, 663419, 2007.

6) M. Ambrossio et al, Nucl. Instr. And Meth, A589,398, 2008.

7) F.Rao et al, Nanotechnology, 20, 055501, 2009.

8) L.X. Zheng et al, Nature Materials, 3, 673, 2004.

9) P.Tenenbaum, Ts. Shintake, Ann. Rev. Nucl. Part. Sc. 49, 125, 1999; SLAC-PUB 8057.

10) E. Bravin, Proc. DIPAC 2007, Venice, Italy, MOO1A02, p.16.

11) http://www.pa.msu.edu/cmp/csc/ntproperties (Compendium compiled by T.A. Adams II).

12) J.-P. Salvetat et al, Appl. Phys. Lett. A69, 255, 1999.

13) Ts. Shintake, Nucl. Instr. and Meth. A311, 453, 1992.

14) K.A. Ispirian, M.K. Ispirian and R.K. Ispirian, Nucl. Instr. And Meth. A336, 423, 1993.

15) S.G. Arutunian, M.R. Mailian, Kay Wittenburg, Nucl. Instr. and Meth. A572, 1022, 2007. S.G. Arutunian, «Vibrating wire sensors for beam instrumentation», To be Published.

16) G.Charpak, Ann. Rev. Nucl. Sc. 20, 195. 1970.

17) F. Sauli, Preprint CERN-77-09, 1977.

18) N. Prakash M., Dissertation, Florida State University, USA, 2005.

19) X.W. Tang et al, Phys. Med. Biol., 50, 23-25, 2005.

20) M.F. Yu et al. Proc. of 6-th Foresight Conf. on Molecular Nanotechnology, Nov. 12-15,1998, Santa Clara, CA,USA, published in a special issue of Nanotechnology.

21) H.Liu et al, Nanotechnology, 19, 445716, 2008.

22) S.K. Esin et al, Proc. Linac96,Geneva, Switzerland, 1996, p. 193.

23) A.V. Okatrub et al, Pisma Zh. Eksp. Toer. Fiz. 81, 37, 2005.

24) P. Elmfors et al, Nucl. Instr. and Meth. A396, 13, 1997.

25) K.A. Ispirian, A.T. Margarian, A.M. Zverev, Nucl. Instr. and Meth. 117, 125, 1974.

26) L.D. Landau, J. Phys. USSR, 8, 201, 1944.

27) O. Blunck, S. Leisegang, Z. Phys. 128, 500, 1950.

28) K.A. Ispirian, A.T. Margarian, Nucl. Instr. and Meth. B36, 364, 1989.

29) H. Kolbenstvedt, J. Appl. Phys.. 38, 4785, 1967; 46, 2771, 1975.

30) S.M. Darbinian, K.A. Ispirian, Radiation Effects, 62, 207, 1982; Preprint YerPhI, 461-3-81, 1981.

31) L.M. Middleman, R.L. Ford, R. Hofstadter, Phys. Rev. 2A, 1429, 1970.

32) M.E. Rose, S.A. Korff, Phys. Rev. 59, 850, 1941.

33) W. Diethorn, NYO-6628, 1956; see R.W. Kiser, Appl. Scient. Res. B8, 183, 1960.

34) A. Williams, R.I. Sara, Int. J. Appl. Rad. Isotopes, 13, 229, 1962.

35) W. Akande, Rev. Sc. Instr. 63, 4354, 1992.

36) G.E. Kocharov, G.A. Korolev, Izv. Akad. Nauk SSSR, 27, 301, 1963.

37) H. Date, M. Shimozuma, Phys. Rev. E64, 066410, 2001.

38) P.J.B.M. Rachinhas et al, IEEE, NS, 1, 90, 1995.

39) H. Loyola, L. Birstein, A. Hevia, Radiation Protection Dosimetry, 132(1):25, 2008.

40) P. Avouris, Physics Today, 62, 34, 2009.